# FedCL-Ensemble Learning: A Framework of Federated Continual Learning with Ensemble Transfer Learning Enhanced for Alzheimer's MRI Classifications while Preserving Privacy

Rishit Kapoor
School of Computer Science and Engineering, Vellore Institute of Technology, Vandalur Kelambakkam Road, Chennai 600127, India
rishit.kapoor2021@vitstudent.ac.in

Jesher Joshua
School of Computer Science and Engineering, Vellore Institute of Technology, Vandalur Kelambakkam Road, Chennai 600127, India
jesherjoshua.m2021@vitstudent.ac.in

Muralidharan Vijayarangan
School of Computer Science and Engineering, Vellore Institute of Technology, Vandalur Kelambakkam Road, Chennai 600127, India
muralidharan.v2021@vitstudent.ac.in

Natarajan B
School of Computer Science and Engineering, Vellore Institute of Technology, Vandalur Kelambakkam Road, Chennai 600127, India
rec.natarajan@gmail.com

*Abstract*— This research work introduces a novel approach to the classification of Alzheimer's disease by using the advanced deep learning techniques combined with secure data processing methods. This research work primary uses transfer learning models such as ResNet, ImageNet, and VNet to extract high-level features from medical image data. Thereafter, these pre-trained models were fine-tuned for Alzheimer's related subtle patterns such that the model is capable of robust feature extraction over varying data sources. Further, the federated learning approaches were incorporated to tackle a few other challenges related to classification, aimed to provide better prediction performance and protect data privacy. The proposed model was built using federated learning without sharing sensitive patient data. This way, the decentralized model benefits from the large and diversified dataset that it is trained upon while ensuring confidentiality. The cipher-based encryption mechanism is added that allows us to secure the transportation of data and further ensure the privacy and integrity of patient information throughout training and classification. The results of the experiments not only help to improve the accuracy of the classification of Alzheimer's but at the same time provides a framework for secure and collaborative analysis of health care data.

## I. Introduction

Alzheimer's disease, a progressive neurodegenerative disorder, has become a notable challenge of modern healthcare and has adversely affected millions of people across the globe [1]. Early and accurate diagnosis is vital for successful treatment and effective planning. Recently, Machine Learning (ML) and Deep Learning (DL) techniques have been able to present promising results on medical imaging data, especially MRI, for these modalities for the detection and classification of different stages of Alzheimer's disease [2,3].

Therefore, the advanced techniques aforementioned face a few challenges in their transit to practical healthcare applications. First, due to the sensitive nature of medical data, serious privacy issues arise, which make aggregation in large-scale datasets inappropriate and infeasible, and this further restricts the possibility of robust model training [4]. Moreover, data heterogeneity among medical institutions and the ever-changing nature of the disease necessitates models capable of continuous adaptation and learning from new data without eroding performance on old tasks [5,6].

To counteract this challenge, this paper proposes FedCL-Ensemble, a novel framework that amalgamates Federated Learning, Continual Learning, and Ensemble Transfer Learning techniques for privacy-preserving Alzheimer's MRI classification. In this case, the research work aims to classify the MRI scans into four categories: Non-Demented, Very Mild Demented, Mild Demented, and Moderate Demented; thus, offering an integrative evaluation of disease progress.

The heart of this framework makes use of transfer learning through pre-trained models such as Inception V3 to extract high-level features from MRI scans [7]. These models that are initially trained on large image datasets are further fine-tuned to retrieve subtle features that could characterize the various stages of Alzheimer's disease. They are made use of in an ensemble model to achieve better performance and generalizability in the setting of classification [8].

To allow preservation of data privacy and collaborative learning across multiple institutions, this paper uses a federated learning framework [9,10]. Such arrangements allow contributing healthcare providers to train models on their data sets without ever having to share raw patient information, eliminating dire privacy concerns associated with medical data. Furthermore, this paper offers up a very novel encryption mechanism allowing for the training of the classification network on ciphertexts rather than original images, creating an extra layer of security assuring protection against potential data breaches while in transit or during deployment of the trained models [11].

At the node level, this paper explores continual learning strategies to allow the model to adapt to new data distributions and changing disease patterns as they evolve over time [12,13].

This would serve to alleviate catastrophic forgetting a well-known issue with neural networks-whereby the model loses performance on certain classes it had previously learned while accumulating new information.

The research study contributes to persistent technical advancement concerning the security and adaptability of machine-learning models for medical imaging while striking at the very chord of the need for credible privacy-preserving collaborative approaches in healthcare AI [14,15]. By enabling those mechanisms, we have built the most robust and generalizable models without compromising on patient privacy, and ethical adoption of AI in the diagnosis and therapeutic monitoring of Alzheimer's disease.

In the next sections, this paper describes the architecture of FedCL-Ensemble, experimental setups, and results. It shall also provide implications of findings with regard to future research and clinical applications in the management of Alzheimer's disease alongside a larger realm of medical image analysis.

## II. RELATED WORKS

The research intersects several key areas in machine learning and medical image analysis. In their review, L. Geert et al. [12] have brought to light most recent applications of deep learning in medical imaging, particularly emphasizing rapid adoption and applications of convolutional networks. In particular, these networks are even proving successful in relatively difficult tasks such as image classification/detection and segmentation. The research work incorporates these principles, particularly dealing with MRI scan classification in facilitating dementia detection. It has been shown by S. M. McKinney et al. [1] that AI systems have surpassed human experts in the analysis of breast cancer in medical image data for screening purposes, illustrating an impressive deep learning model functioning solution of false positives and false negatives; thus, in so doing, the research strives to enhance the classification of MRI scans for dementia diagnosis. Transfer learning has been described as the most popular technique to accomplish medical image analysis, which allows knowledge from pre-trained networks. Z. Fuzhen et al. [4] provided a complete survey of transfer learning techniques, enunciating applications in diverse domains, along with the healthcare domain. This research uses transfer learning for better classification of MRI scans through the model Inception V3.

Federated learning emerged as an important privacy-preserving tool for collaborative training of models. The research work of L. Qinbin et al. [2] in federated learning finds its application in the critical areas like healthcare, where data privateness is of utmost importance. Therefore, this paper employees federated learning to allow model training in collaboration between various institutions without compromising patient privacy. G. Hao et al. [16] gave a comprehensive survey on the federated learning methods specifically in the field of medical image analysis. They classify existing methods based on client-side methods, server-side methods, and communication methods, allowing insight into the motivation behind various approaches. This paper adopts the federated learning framework for dementia classification from MRI scans, which places the research work within that tradition. B. Pfitzner et al. [19] conducted a systematic literature review of federated learning in medical contexts, where they argue the necessity for preserving patient anonymity while enabling collaborative research. Their core statement highlights the sizable potential of this direction on combining data from several sites throughout the world while still maintaining data on patients' privacy-centricity-very much at the focus of the proposed approach.

In common parlance, continuous learning is conceived as a balance between developing models that consume new information while avoiding recall errors on the previously-acquired knowledge. A. R. Andrei et al. [5] proposed progressive neural networks, an approach to solve complex sequences of tasks with transfer knowledge while avoiding catastrophic forgetting. K. James et al. [6] have sought to neutralize catastrophic forgetting in neural networks by controlling the learning surface with an increased curvature on neurons strongly correlated with previous tasks.

Recent works have examined the merge of Federated Learning and continual learning to handle adaptive and privacy-conserving learning challenges in distributed frameworks. W. Liyuan et al. [18] survey provides a systematic overview of continual learning. They delve at length into theoretical foundations, methods, and applications. Emphasis is placed on achieving an efficient stability-plasticity trade-off along with the necessary intra/inter-task generalizability in resource-constrained environments. A proposal by J.M. Jesher et al [20]. presents a paradigm combining federated learning and continual learning for web phishing detection. This enables the conversion of a federated model from one or more distributed nodes to assess both individuals and generalization without storing data, followed by performing federated aggregation at a central server. This paper presents good evidence that the merger of federated and continual learning indicates adaptive and privacy-protective model training. As for the burden imposed by privacy on deep learning, a number of approaches have come up for the protection of sensitive information. R. Shokri and V. Shmatikov [13] have presented a differentially private deep learning system that allows multiple parties to collaboratively and privately train an accurate neural network model without sharing their respective input datasets. Algorithmic ways of learning with differential privacy were developed by M. Abadi et al. [14], showing that this paper can indeed train deep neural networks with differential privacy under a reasonable privacy budget. Our work in the paper here extends the privacy-preserving methodology by offering a cryptographic tool for encrypting input images before processing the images on the Inception V3 model. This approach adds another layer of protection to privacy in the federated learning framework itself. In the context of multi-center collaborative learning, E. Zhu et al. [21] presented MP-Net, an approach securing image segmentation in collaborative medical applications against data corruptions. It mainly provides a new avenue in which data transmission can be reduced while its privacy can be safeguarded in the collaborative learning scenario. While the implementation differs in the ensuing stages of this paper, nevertheless it is aligned in promoting a secure and analyzed multi-center medical collaboration based on imaging expertise.

Finally, the present research work builds upon and extends beyond the diverse strands of research by incorporating transfer learning, federated learning, continual learning, and privacy-preserving techniques to develop a secure, adaptive framework for MRI-based dementia classification. These approaches are integrated so that attempts can be made to solve some realistic issues in medical image analyses such as data privacy, model adaptability, and collaborative learning.

### III. DATASET

This research uses MRI scans as input data to extract structural information that is important in differentiating phases of Alzheimer's disease progression. Class representation diversity in the dataset addresses the initial objective of building a robust classification model. The dataset used in this study consists of MRI images, classified into four different classes: Non-Demented, Very Mild Demented, Mild Demented, and Moderate Demented. This dataset is sourced from Kaggle and has been curated from multiple online sources. Each label was checked manually for assurance of accuracy. The images are then split into training and testing subsets to ensure a fair representation of each class for model evaluations.

| Class | Count (Original) | Count (Balanced) |
|---|---|---|
| Non-Demented | 3200 | 3200 |
| Very Mild Demented | 2240 | 3200 |
| Mild Demented | 896 | 3200 |
| Moderate Demented | 64 | 3200 |

Distribution of the dataset was extremely imbalanced: 3200 images in the Non-Demented class, 2240 images in the Very Mild Demented class, 896 images in the Mild Demented class, and only 64 images in the Moderate Demented class. The response to this imbalance and the resulting model learning bias to the majority class were addressed through Synthetic Minority Oversampling Technique (SMOTE) to upsample the minority classes. After the application of SMOTE, the dataset is balanced so that each class has 3200 images—thereby allowing for equal representation; thereby there is no hindrance for model learning across all the Alzheimer type stages; in addition, the FedCL-Ensemble framework worked while maintaining privacy through federated learning.

### IV. METHODOLOGY

FedCL-Ensemble, a Federated Continual Learning framework that integrates ensemble transfer learning and privacy-preserving techniques for MRI-based Alzheimer's disease classification, is proposed in this study. The methodology is fairly detailed, comprising stages such as dataset preprocessing, model selection, federation learning, and continual learning strategies.

### V. DATA PREPROCESSING

Standard medical image processing methods were used to prepare the MRI images for effective feature extraction from the dataset and enhance their quality. The images were resized to 176×176 pixel dimensions and filtered using Gaussian smoothing usually done for the reduction of noise and contrast enhancement for highlighting the critical features. This preprocessing thus made sure that the input data were normalized and uniform across all classes before going for training through the deep learning model.

### VI. MODEL SELECTION

To classify Alzheimer's disease, this paper analyzed and compared five most prominent pretrained models: VGG16, ResNet50, InceptionV3, MobileNetV2, and Xception the training graphs of which are showcased in Fig1. The models were initialized with the weights pretrained on the ImageNet dataset and modified to dismiss the top layers; i.e., the classification layers into the Alzheimer's MRI dataset. The high-level features of the MRI images were then pulled from these models and put into a classification evaluation. Among them, InceptionV3 performed best, giving better accuracy and generalization through the different classes. Therefore, InceptionV3 becomes the best choice of a baseline model to carry out further experiments.

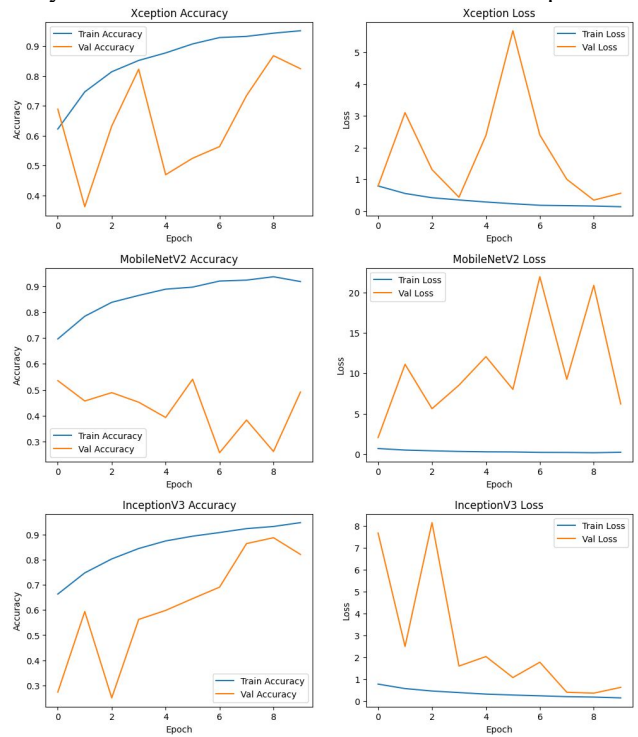

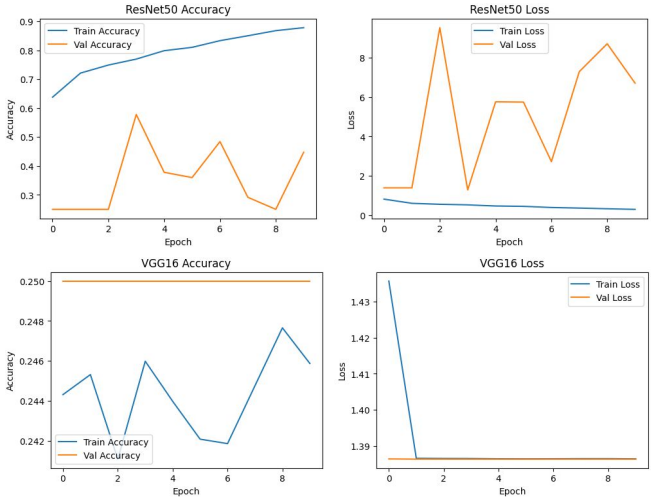

Fig. 1. Accuracy and loss training graphs of VGG16, ResNet50, InceptionV3, MobileNetV2, and Xception models.

## VII. FEDERATED LEARNING SETUP

To mitigate the concerns regarding privacy, this paper employed a federated learning (FL) setup. In this, several nodes representing different healthcare institutions participated in the collaborative training process without exchanging their raw data. Each node trained its local model on the MRI data held at that specific institution. The model updates from the individual nodes were then aggregated using the Federated Averaging (FedAvg) algorithm into a central server, where all locally computed updates were combined to produce a global model. To further enhance data security, the research introduces a cipher encryption mechanism. This encryption model was designed using convolutional layers; using this model, the raw MRI images were encrypted in ciphertext. From there, this ciphertext was then passed through the InceptionV3 model at the nodal level, ensuring that no raw medical data existed even when undergoing federated learning. Therefore, this guarantees the privacy of the patient's data throughout training and model transmission.

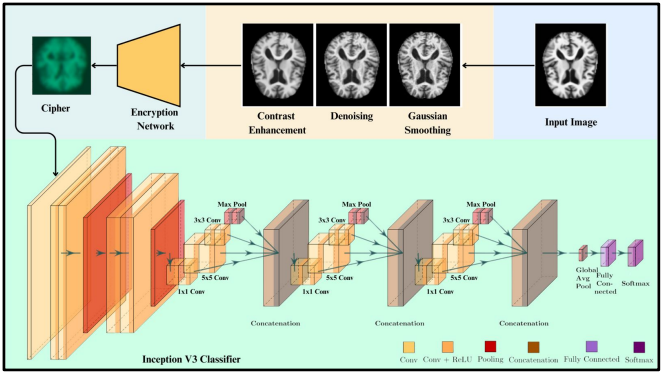

Fig. 2. Inception V3 Classifier and Encryption Network Architectural Setup

## VIII. CONTINUAL LEARNING STRATEGIES

Given the dynamic aspect of Alzheimer's Disease and data heterogeneity across institutions, continual learning (CL) was integrated within the model to learn new data without losing previous knowledge-a process called hindsight learning. Various strategies of continual learning were perpetually compared, including Elastic Weight Consolidation.

This includes other methods such as replay-based techniques, cumulative learning, and Learning without Forgetting (LwF). Here, these strategies would be tested for their strengths in preserving previous model performance in learned tasks using new knowledge from incoming data. These were obviously aimed at imbuing existing high accuracy and robustness capacity in the diagnosis of Alzheimer's disease irrespective of the incremental training approach-the stages of progression of the disease.

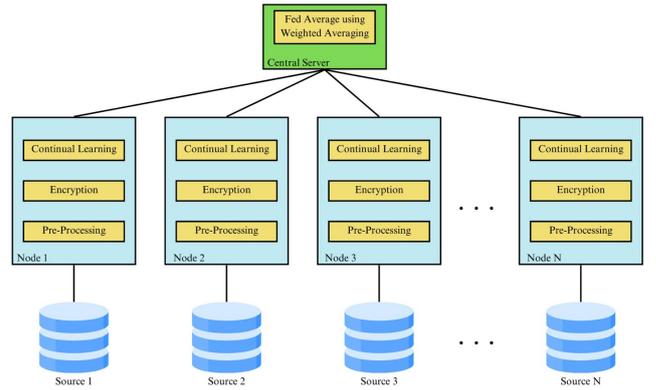

Fig. 3. Continual learning Process

## IX. FEDCL-ENSEMBLING

The aggregation mechanism plays a crucial role in FedCL-Ensemble for combining model updates from several nodes (hospitals or institutions) without sacrificing the ability of continual learning. The FedAvg algorithm is employed to gather the model updates at the central server, whereby continual-learn strategies are utilized at each node to tackle evolving data distributions and tasks with.

Let there be K participating nodes (institutions), with a local model $w_k$ owning the corresponding dataset $D_k$, where k∈{1,2,…,K}.. The main aim is to aggregate the local models and develop a global model $w_G$ such that it minimizes some global objective function that is based on the local models, subject to the privacy and evolution over time to new data distributions. At each communication round t, the nodes update the local model independently based on local data. Then, the aggregate takes place through an update of the central server using a weighted average according to the sizes of the datasets.

$$w_G^{(t+1)} = \sum_{k=1}^{K} \frac{|D_k|}{\sum_{i=1}^{k}|D_i|} w_k^{(t+1)} \qquad (1)$$

where $w_G^{(t+1)}$ is the global model at round (t+1), and $|D_k|$ is the number of samples at node k. The term represents the weight of each local model, proportional to the size of its dataset. This weighted average ensures that nodes with larger datasets have a greater influence on the global model.

To incorporate continual learning strategies, this paper modifies the standard federated learning process to include the capability of preserving previously learned knowledge while adapting to new tasks or data. Specifically, directs each node to run a continual learning strategy, such as Elastic Weight Consolidation (EWC) [1], or replay-based methods, during the local update to ensure that the local model retains performance in previously encountered tasks while new tasks arrive. For example, in the case of EWC, the local objective function (Equation 2) on every federated round is modified by a regularizing term so that it penalizes changes to important weights, thus impeding catastrophic forgetting. The objective function for EWC is defined as:

$$L_k(w_k) = L_{task}(w_k) + \frac{\lambda}{2}\sum_i \Omega_i (w_k - w_{k,prev})^2 \quad (2)$$

where $L_{task}(w_k)$ is the standard task loss on the local dataset, $\Omega_i$ represents the importance of the i-th parameter (based on the Fisher Information Matrix), and $w_{k,prev}$ is the parameter value from previous tasks. This method effectively combines federated learning with continual learning, allowing the model to benefit from diverse data sources and adapt to evolving information

## X. EVALUATION METRICS

The aggregation mechanism plays a crucial role in FedCL-Ensemble for combining model updates from several nodes (hospitals or institutions) without sacrificing the ability of continual learning. The FedAvg algorithm is employed to gather the model updates at the central server, whereby continual-learn strategies are utilized at each node to tackle evolving data distributions and tasks with.

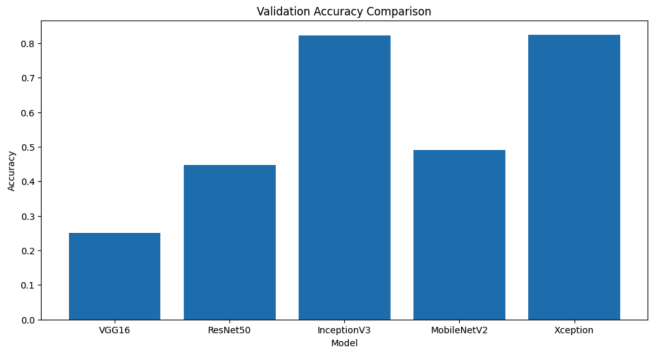

Fig. 4. Validation Accuracy of different CV Vision Models

## XI. RESULTS AND DISCUSSIONS

The FedCL-Ensemble framework was evaluated with InceptionV3, which was trained on the encrypted MRI images to guarantee patient data privacy. The base InceptionV3 network trained on ciphertext for 10 epochs achieved initial accuracy of 73.68%. Model updates were then readied for distribution via a federated learning environment.

The federated learning process ran across 20 rounds of training, with each round having 50 epochs. Five nodes were involved in the training, representing different institutions, with the central server aggregating model updates via the Federated Averaging algorithm. After 20 rounds, the model achieved an amazing accuracy of 97.8%, with the EWC strategy being used to carry out an effective task of catastrophic forgetting resolution.

This paper further evaluated the robustness and adaptability of the model over time, comparing a multitude of different continual learning (CL) strategies: EWC, replay based learning, cumulative learning, and the Learning without Forgetting strategy. The table below summarizes the performance of each CL strategy based on the mentioned standard classification metrics with their standard deviations across a number of runs.

| CL Strategy | Accuracy (%) | Precision (%) | Recall (%) | F1 Score (%) |
|---|---|---|---|---|
| **Elastic Weight Consolidation (EWC)** | **97.8 ± 0.42** | **97.5 ± 0.35** | **97.9 ± 0.41** | **97.7 ± 0.38** |
| Replay | 95.2 ± 0.51 | 94.9 ± 0.48 | 95.3 ± 0.47 | 95.1 ± 0.46 |
| Cumulative Learning | 93.4 ± 0.62 | 93.1 ± 0.60 | 93.5 ± 0.58 | 93.3 ± 0.59 |
| Learning without Forgetting (LwF) | 91.7 ± 0.69 | 91.3 ± 0.66 | 91.8 ± 0.64 | 91.6 ± 0.63 |

The results indicate that EWC has better accuracy, precision, recall, and F1-score compared with each of the other methods. Its ability to retain knowledge learned prior to consecutive collection of data was evidenced to contribute mostly to the great performance of federated learning. The accuracy of 97.8% points to promise since the model claimed to accurately distinguish Alzheimer's disease stages across multiple institutions, assuring data privacy. Replay-based learning, where some subset of the data is fed back into the model while it is being trained, performed competitively in its application, with an accuracy of 95.2%. Cumulative learning and learning without forgetting (LwF) methods, albeit effective, lagged further behind; leading to the suggestion that these methods could not keep up the performance of the models over sequential tasks in a federated setup. It is clear that combining federated learning from continuous learning strategies is a viable means of remedying two grave challenges present in medical imaging: data privacy as well as continual learning. The federated learning approach, coupled with encryption mechanisms, ensured that sensitive patient data were never shared while enabling sound collaborative model training. Overall, the FedCL-Ensemble framework not only boosts the accuracy of Alzheimer's classification but also sets a benchmark for privacy-aid collaborative learning in healthcare. More future work may also investigate other strategies that can be used to enhance the continual learning process and widen the application of the framework into yet other medical fields with similar privacy concerns.

## XII. CONCLUSION

The focus of the research constitutes the FedCL-Ensemble, a framework of federated continual learning designed for the private classification of Alzheimer's disease via MRI images; it integrates advanced deep learning techniques such as transfer ensemble learning, continual learning strategies, and the very interesting federated learning perspective, while ensuring high classification accuracy and privacy of records. The comparisons with different pre-trained models led to identifying InceptionV3 as the most optimal choice for feature extraction in Alzheimer's disease classification. The federated learning algorithm, combined with a cipher Encryption mechanism, enabled the collaboration of several nodes in the training of the model while at the same time protecting the patients' privacy.

The results show how the proposed framework, equipped with the EWC method, attained an astounding percentage of 97.8% that surpasses the alternative continual learning strategies, such as replay-based learning, cumulative learning, and LwF, thus proving the strength and security of FedCL-Ensemble as a solution for privacy-preserving collaborative analysis in healthcare data.

## XIII. FUTURE WORK

Several directions for future investigation emerge from the present study. Further, while EWC demonstrated effectiveness, taking a cue from a meta-learning technique or an orthogonal gradient descent base could enhance performance in a federated learning setup in future research on the same topic. Further exploring alternative medical imaging domains that remain confidentiality-driven for the proposed framework such as cancer detection, diabetic retinopathy classification, and cardiovascular disease prediction would showcase the inherent versatility inherent and its scalability.

Another area under improvement would be to further optimize encryption mechanisms in the learning process. Future research could investigate more lightweight and efficient encryption techniques in order to reduce computational overhead whilst satisfying data security. In addition, the examination of personalized federated learning techniques may prove useful in customizing model performances for a particular institution or patient population, without compromising privacy or generalizability. Finally, the prolonged evaluation of the framework in a continual learning manner with frequent and diverse data updates would allow for such insights as its scalability and robustness in real-world healthcare applications. The way these future directions shape up we will see the FedCL-Ensemble framework morph into a wonderful tool for secure, scalable, adaptive AI-driven diagnosis which will contribute to better patient care and outcomes.